\documentclass[aps,prb,reprint,amsmath,amssymb,superscriptaddress]{revtex4-2}
 
\usepackage{amsmath}
\usepackage{amssymb}
\usepackage{graphicx}
\usepackage{CJK}
\usepackage{keyval}
\usepackage{dcolumn}
\usepackage{bm}
\usepackage{color}
\usepackage{txfonts}
\usepackage{verbatim}
\usepackage[rightcaption]{sidecap}

\usepackage{soul}
\usepackage{ulem}
\usepackage{booktabs}
\usepackage{lineno}
\usepackage{sidecap}
\usepackage[colorlinks=true, citecolor=blue, linkcolor=blue, urlcolor=blue]{hyperref}

\begin{document}

\title{$3d_{z^2}$ orbital delocalization and magnetic collapse in superconducting (La,Pr)$_3$Ni$_2$O$_{7-\delta}$ films}

\author{Xiaoyang Chen}
\thanks{Equal contribution}
\affiliation{Diamond Light Source, Harwell Campus, Didcot, OX11 0DE, United Kingdom}

\author{Wenliang Zhang}
\affiliation{National Synchrotron Radiation Laboratory and School of Nuclear Science and Technology, University of Science and Technology of China, Hefei 230026, China}

\author{Fei Peng}
\thanks{Equal contribution}
\affiliation{State Key Laboratory of Quantum Functional Materials, Department of Physics, Guangdong Basic Research Center of Excellence for Quantum Science, Southern University of Science and Technology, Shenzhen, China}

\author{Ting Cui}
\thanks{Equal contribution}
\affiliation{Beijing National Laboratory for Condensed Matter Physics and Institute of Physics, Chinese Academy of Sciences, Beijing, China}
\affiliation{Department of Physics \& Center of Materials Science and Optoelectronics Engineering, University of Chinese Academy of Sciences, Beijing, China}

\author{Guangdi Zhou}
\thanks{Equal contribution}
\affiliation{State Key Laboratory of Quantum Functional Materials, Department of Physics, Guangdong Basic Research Center of Excellence for Quantum Science, Southern University of Science and Technology, Shenzhen, China}
\affiliation{Quantum Science Center of Guangdong-Hong Kong-Macao Greater Bay Area, Shenzhen, China}

\author{Zezhong Li}
\affiliation{National Synchrotron Radiation Laboratory and School of Nuclear Science and Technology, University of Science and Technology of China, Hefei 230026, China}

\author{Jaewon Choi}
\affiliation{Diamond Light Source, Harwell Campus, Didcot, OX11 0DE, United Kingdom}

\author{Lizhi Xu}
\affiliation{State Key Laboratory of Quantum Functional Materials, Department of Physics, Guangdong Basic Research Center of Excellence for Quantum Science, Southern University of Science and Technology, Shenzhen, China}

\author{Yiu-Fung Chiu}
\affiliation{Diamond Light Source, Harwell Campus, Didcot, OX11 0DE, United Kingdom}
\affiliation{Department of Physics, University of Oxford, Clarendon Laboratory, Parks Road, Oxford OX1 3PU, United Kingdom}

\author{Stefano Agrestini}
\affiliation{Diamond Light Source, Harwell Campus, Didcot, OX11 0DE, United Kingdom}

\author{Sahil Tippireddy}
\affiliation{Diamond Light Source, Harwell Campus, Didcot, OX11 0DE, United Kingdom}

\author{Haoliang Huang}
\affiliation{State Key Laboratory of Quantum Functional Materials, Department of Physics, Guangdong Basic Research Center of Excellence for Quantum Science, Southern University of Science and Technology, Shenzhen, China}
\affiliation{Quantum Science Center of Guangdong-Hong Kong-Macao Greater Bay Area, Shenzhen, China}

\author{Heng Wang}
\affiliation{State Key Laboratory of Quantum Functional Materials, Department of Physics, Guangdong Basic Research Center of Excellence for Quantum Science, Southern University of Science and Technology, Shenzhen, China}
\affiliation{Quantum Science Center of Guangdong-Hong Kong-Macao Greater Bay Area, Shenzhen, China}

\author{Xianfeng Wu}
\affiliation{State Key Laboratory of Quantum Functional Materials, Department of Physics, Guangdong Basic Research Center of Excellence for Quantum Science, Southern University of Science and Technology, Shenzhen, China}

\author{Peng Li}
\affiliation{State Key Laboratory of Quantum Functional Materials, Department of Physics, Guangdong Basic Research Center of Excellence for Quantum Science, Southern University of Science and Technology, Shenzhen, China}
\affiliation{Quantum Science Center of Guangdong-Hong Kong-Macao Greater Bay Area, Shenzhen, China}

\author{Jin-Feng Jia}
\affiliation{State Key Laboratory of Quantum Functional Materials, Department of Physics, Guangdong Basic Research Center of Excellence for Quantum Science, Southern University of Science and Technology, Shenzhen, China}
\affiliation{Quantum Science Center of Guangdong-Hong Kong-Macao Greater Bay Area, Shenzhen, China}

\author{Mirian Garcia-Fernandez}
\affiliation{Diamond Light Source, Harwell Campus, Didcot, OX11 0DE, United Kingdom}

\author{Yi Lu}
\affiliation{National Laboratory of Solid State Microstructures and Department of Physics, Nanjing University, Nanjing 210093, China}
\affiliation{Collaborative Innovation Center of Advanced Microstructures, Nanjing University, Nanjing 210093, China}

\author{Er-Jia Guo}
\thanks{Corresponding author: ejguo@iphy.ac.cn}
\affiliation{Beijing National Laboratory for Condensed Matter Physics and Institute of Physics, Chinese Academy of Sciences, Beijing, China}
\affiliation{Department of Physics \& Center of Materials Science and Optoelectronics Engineering, University of Chinese Academy of Sciences, Beijing, China}

\author{Qi-Kun Xue}
\affiliation{State Key Laboratory of Quantum Functional Materials, Department of Physics, Guangdong Basic Research Center of Excellence for Quantum Science, Southern University of Science and Technology, Shenzhen, China}
\affiliation{Quantum Science Center of Guangdong-Hong Kong-Macao Greater Bay Area, Shenzhen, China}

\author{Zhuoyu Chen}
\thanks{Corresponding author: chenzhuoyu@sustech.edu.cn}
\affiliation{State Key Laboratory of Quantum Functional Materials, Department of Physics, Guangdong Basic Research Center of Excellence for Quantum Science, Southern University of Science and Technology, Shenzhen, China}
\affiliation{Quantum Science Center of Guangdong-Hong Kong-Macao Greater Bay Area, Shenzhen, China}

\author{Donglai Feng}
\thanks{Corresponding author: dlfeng@ustc.edu.cn}
\affiliation{New Cornerstone Laboratory, Hefei National Laboratory, Hefei, 230088, China}

\author{Ke-Jin Zhou}
\thanks{Corresponding author: kjzhou@ustc.edu.cn}
\affiliation{National Synchrotron Radiation Laboratory and School of Nuclear Science and Technology, University of Science and Technology of China, Hefei 230026, China}

\begin{abstract}
The recent discovery of Ruddlesden–Popper (RP) nickelate thin-film superconductors has opened a new frontier in unconventional superconductivity. Its realization requires both compressive epitaxial strain and highly oxidative growth conditions, yet the microscopic pathway from the parent phase to the superconducting phase remains elusive. Here, X-ray absorption spectra and resonant inelastic X-ray scattering are employed to track this evolution by independently tuning strain and oxygen content in (La,Pr)$_3$Ni$_2$O$_{7-\delta}$ thin films. We uncover a remarkable two-step narrative. First, signatures of delocalization emerge in the same way upon two independent tunings: Spectral weight transfers from a "Upper Hubbard"-like peak to the hole-like peak associated with O $2p_z$ state, and in parallel, the initially localized Ni $3d_{z^2}$ orbital becomes more itinerant followed by the broadening and weakening of $dd$ orbital excitations. Second, as itinerancy increases, long-range spin-density-wave (SDW) order is suppressed in both intensity and correlation length, indicating direct competition with superconductivity. Yet, short-range magnons persist: they become damped but their bandwidth stays unchanged. Our results paint a coherent picture that both strain and oxygenation drive the RP bilayer nickelates towards the superconducting instability, where the O $2p_z$ and Ni $3d_{z^2}$ orbitals become delocalized. Concomitantly, the long-range magnetic order loses coherence and gets suppressed. These findings establish an orbital-selective route to RP nickelate superconductivity, in which the delocalization of the $2p_z$ and $3d_{z^2}$ orbitals and the robust short-range magnons upon the melting of SDW order are prerequisites, providing strong constraints for theory and the roadmap for designing nickelate superconductors.

\end{abstract}

\date{\today}

\maketitle

\vspace{8mm}

\section{Introduction}
The Ruddlesden--Popper (RP) nickelates ($R_{n+1}\mathrm{Ni}_n\mathrm{O}_{3n+1}$) have emerged as a new family of high-temperature superconductors. Structurally, they consist of $n$ layers of perovskite-like $\mathrm{NiO}_6$ octahedra stacked along the $c$ axis and separated by rock-salt $\mathrm{LaO}$ layers, with adjacent octahedra sharing apical oxygens. Superconductivity is realized under hydrostatic pressure, with the superconducting transition temperature ($T_\mathrm{c}$) tuned by the dimensionality ($i.e.$, the layer number $n$), or, the stacking sequence ~\cite{sun2023signatures,Zhang2024high-temperature,Wang2024bulk,zhu2024superconductivity,shi2025pressure,Li2026bulk}. Among this family, the Sm-doped bilayer RP nickelate exhibits the highest reported $T_\mathrm{c}$, reaching $\sim 96$~K under pressure~\cite{Li2026bulk}. In contrast to cuprate superconductors, both Ni $d_{x^2-y^2}$ and $d_{z^2}$ orbitals contribute to the low-energy electronic structure~\cite{Chen2024orbital,chen2024electronic}, giving rise to multi-orbital characteristics reminiscent of iron-based superconductors~\cite{stewart2011superconductivity}. A variety of theoretical models have been proposed to describe its electronic structure and the pairing symmetry under pressure, with ongoing debate regarding the relative importance of the partially filled Ni $d_{z^2}$ and $d_{x^2-y^2}$ orbitals in driving superconductivity~\cite{Wang2025recent,Puphal2026superconductivity}. Besides, the evolution of the magnetic properties such as the spin-density-wave (SDW) is critical to identify the pairing mechanism \cite{chen2025charge}. However, direct experimental verification remains challenging due to the extreme conditions imposed by high pressure.

Recent advances in thin-film growth techniques have enabled the synthesis of RP and hybrid RP nickelate films on various substrates, introducing tunable biaxial strain via lattice mismatch. Superconductivity has been realized in compressively strained (La,Pr)$_3$Ni$_2$O$_{7-\delta}$ films and related superstructures, with $T_\mathrm{c}^\mathrm{max} \approx 60$~K on SrLaAlO$_4$ (SLAO, $\sim -2\%$ strain) and $\sim 12$~K on LaAlO$_3$ (LAO, $\sim -1\%$ strain)~\cite{ko2025signatures,zhou2025ambient,Liu2025superconductivity,Hao2025superconductivity,SLAO60K,tarn2026reducing,Nie2026superconductivity}, providing an accessible route to investigate the superconducting mechanism at ambient pressure. Structurally, compressive strain reduces the in-plane lattice constant and suppresses the tilting of the $\mathrm{NiO}_6$ octahedra, driving the system toward a higher-symmetry configuration \cite{bhatt2026structural}. 

Besides the structural change, the strain also leads to the modifications in the electronic and magnetic properties. Density functional theory (DFT) calculations predict that the $d_{z^2}$-derived bands shift away from the Fermi level ($E_\mathrm{F}$), though under hydrostatic pressure they may cross $E_\mathrm{F}$~\cite{yi2025unifying}. Experimentally, it has been an important question for angle-resolved photoemission spectroscopy (ARPES) studies regarding the position and role of the $d_{z^2}$-derived bands in compressively strained films \cite{Nie2026superconductivity,wang2025electronic,LPARPES,Shen2025nodeless,Sun2025observation,Li2026threedimensional}. 
Concerning the spin degree of freedom, the interlayer exchange coupling ($J_\perp$) is predicted to be enhanced owing to the reduced tilting of NiO$_6$ octahedra under compressive strain, which may play a key role in stabilizing superconductivity, a scenario supported by recent resonant inelastic X-ray scattering (RIXS) studies~\cite{zhong2025spin}. Optical spectroscopy revealed the evolution of a density-wave-like gap in La$_3$Ni$_2$O$_{7-\delta}$ under pressure which suggests a potential competition between density-wave order and superconductivity~\cite{meng2024density}. However, the direct experimental probe of such density-wave order evolution in compressively strained thin films remains unexplored.

From a materials perspective, oxygen deficiency has been identified as a critical challenge for superconductivity for RP bilayer nickelates \cite{Dong2024visualization,Hao2025superconductivity}. 
These vacancies are detrimental to superconductivity because they may destroy the Ni 3$d_{z^2}$ interlayer coupling and in-plane phase coherence \cite{Wang2025SIT,Wang2026superconducting,Liu2026half-dome}.
Beyond vacancies, interstitial oxygen atoms by extra oxygenation may induce hole carriers that suppress superconductivity \cite{dong2025interstitial,Liu2026half-dome}. It is thus pivotal to experimentally explore the impact of oxygen content on the electronic and magnetic properties. 

In this work, we track the microscopic evolution in (La,Pr)$_3$Ni$_2$O$_{7-\delta}$ thin films from a non-superconducting parent phase to a superconducting state using X-ray absorption spectra (XAS) and resonant inelastic X-ray scattering, where compressive strain and oxygen content serve as two parallel tuning knobs. Extracting from the O $K$-XAS data, we find that the O $2p_z$ and the Ni $3d_{z^2}$ orbitals become more delocalized upon tuning of the strain or oxygen content. The Ni $L$- XAS and RIXS results exhibit that the XAS excitation and $dd$ orbital excitations associated with Ni $3d_{z^2}$ orbital character become broadened and less pronounced despite a rigid crystal-field splitting upon the two independent tunings. Accompanying this orbital-selective delocalization, long-range spin-density-wave order is suppressed in both intensity and correlation length, in a direct competition with superconductivity. Short-range magnons become damped but retain their bandwidth, challenging predictions of enhanced interlayer exchange coupling. Our results establish an orbital-selective route to superconductivity in (La,Pr)$_3$Ni$_2$O$_{7-\delta}$ where the delocalization of the $2p_z$ and $3d_{z^2}$ orbitals and the robust short-range magnons are prerequisites that provide strong constraints for theoretical studies and designing nickelate superconductors.

~\\

\section{Results}
\subsection{Engineering of (La,Pr)$_3$Ni$_2$O$_{7-\delta}$ thin films}

We designed two series of thin films to investigate the effects of compressive strain and oxygen stoichiometry independently (see growth details in Method). The strain-tuning series includes two non-superconducting La$_3$Ni$_2$O$_{7-\delta}$ thin films grown on (001)-oriented LaAlO$_3$ and SrLaAlO$_4$ substrates (non-SC LNO/LAO and non-SC LNO/SLAO1), which impose biaxial in-plane compressive strains of approximately $-1\%$ and $-2\%$, respectively (Fig.~1a). These films are identical to those reported in Ref.~\cite{cui2024strain}. They were grown under the same oxygen partial pressure of $4\times10^{-1}$ mbar therefore viewed to contain a comparable oxygen content. Both samples exhibit insulating behavior at low temperatures~\cite{cui2024strain}. The epitaxial strain is effectively transferred to the thin films, leading to systematic structural modifications~\cite{cui2024strain}. Furthermore, the $\mathrm{NiO}_6$ octahedra become more straightened compared to the La$_3$Ni$_2$O$_7$ bulk single crystal (LNO bulk) at ambient pressure (see details in the Supplementary information). 

The oxygen content–tuning series includes two (La,Pr)$_3$Ni$_2$O$_{7-\delta}$ thin films grown on the SrLaAlO$_4$ substrate \cite{Zhou2025gigantic,zhou2025ambient,SLAO60K}. Both samples were grown under an ozone–oxygen atmosphere and capped with a SrTiO$_3$ layer \cite{Shi2026critical} (Fig. 1b). Oxygen content is controlled $in~ situ$, such that one sample is insulating at low temperatures (non-SC LPNO/SLAO2), whereas the other shows a superconducting transition at approximately 50~K (SC LPNO/SLAO3) (see details in Supplementary information).

In the following sections, we discuss the XAS and RIXS results to track the evolution of the electronic and magnetic properties of (La,Pr)$_3$Ni$_2$O$_{7-\delta}$ under the two independent tuning routes, $i.e.$, the compressive strain and the oxygen content.

\begin{figure}
\centerline{\includegraphics[width=86mm,angle=0]{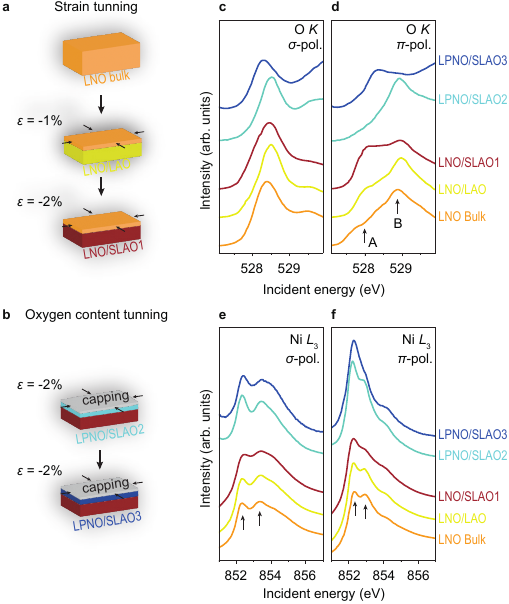}}
\caption{
\textbf{The O $K$- and Ni $L_3$- X-ray absorption spectra.} 
\textbf{a,b,} Schematic illustration of (La,Pr)$_3$Ni$_2$O$_{7-\delta}$ films designed with varying compressive strain (a) and oxygen content (b). The LPNO/SLAO3 is superconducting at low temperature while the other samples are insulating.
\textbf{c,d,} O $K$-XAS measured at $\sigma$ polarisation (c) and $\pi$ polarisation (d). 
\textbf{e,f,} Ni $L_3$-XAS measured at $\sigma$ polarisation (e) and $\pi$ polarisation (f). XAS were measured in total fluorescence yield (TFY) mode at 20~K with a grazing incidence angle of 20$^\circ$, where $\pi$ polarization predominantly probes out-of-plane orbital states, while $\sigma$ polarization probes in-plane orbital states.
}
\label{fig1_xas}
\end{figure}

\subsection{Electronic excitations}


Figures 1c and 1d show the grazing-incident Oxygen $K$-edge X-ray absorption (XAS) spectra from the thin films and LNO bulk reference (See XAS experiments in Methods and the experimental geometry in Supplementary information). For the in-plane direction ($\sigma$-pol), a single-peak profile dominates the pre-edge dictating the O $2p_{x,y}$ orbital character. No appreciable difference is observed in the strain-dependent films in comparison to the LNO bulk. The oxygen content dependent films feature the same pre-edge profile as the LNO bulk, although the resonance of the SC LPNO/SLAO3 has a slight red-shift by $\sim 0.2$~eV. This is due to the higher hole-doping level in the SC LPNO/SLAO3 comparing to the oxygen-deficient non-SC LPNO/SLAO2, consistent with the hole-doping behaviour of La$_{2-x}$Sr$_x$CuO$_4$ cuprate \cite{chen1992out}. Contrary to the in-plane O $K$-XAS, the out-of-plane pre-edge ($\pi$-pol), characterized by a two-peak structure, manifests a much greater evolution where the spectral weight transfers from the second peak (peak B) to the first peak (peak A) in both sample series, particularly the oxygen content dependence. In La$_{2-x}$Sr$_x$CuO$_4$ cuprate, the high-energy peak is related to the "upper Hubbard band" resulting from a
$3d^9 \rightarrow \underline{1s}3d^{10}$ transition of a rather undoped material; the low-energy peak stems from doping-induced holes resulting mainly from $3d^9\underline{L} \rightarrow \underline{1s}3d^{9}$ transition. Here $\underline{L}$ and  $\underline{1s}$ denote the O $2p$ ligand hole and the O $1s$ core hole, respectively. In our (La,Pr)$_3$Ni$_2$O$_{7-\delta}$ thin films, the spectral weight transfer from the ``upper Hubbard band"-like peak B to the hole-like peak A is reminiscent of the evolution of the O $K$-XAS pre-edge in La$_{2-x}$Sr$_x$CuO$_4$ as a function of Sr doping \cite{chen1991electronic}. The same trend in the strain- and the oxygen content series unambiguously unveils that the two tuning methods work effectively and drive La$_3$Ni$_2$O$_{7-\delta}$ from a Mott-like insulating state towards a delocalized superconducting state.
In contrast to La$_{2-x}$Sr$_x$CuO$_4$ where the spectral weight transfer is realized in both O $2p_{x,y}$ and $2p_{z}$ orbitals, here in (La,Pr)$_3$Ni$_2$O$_{7-\delta}$ it merely happens in the O $2p_{z}$ orbital states suggesting that the apical oxygen is much more susceptible upon tuning underscoring its critical role in superconductivity. 

Figures 1e and 1f show the Ni $L_3$-edge XAS spectra from the thin films and the LNO bulk. The Ni $L_3$ XAS spectra of the LNO bulk exhibits characteristic peaks at $\sim$ 852.3 eV and $\sim$ 853.3 eV for $\sigma$-pol (with the latter at $\sim$ 853.0 eV for $\pi$-pol), corresponding to the main $2p^53d^8$ and the satellite $2p^53d^8\underline{L}$ configuration, respectively~\cite{chen2024electronic}. The two-peak XAS profile is persistent in both series with little change in the excitation to the Ni $3d_{x^2-y^2}$ states ($\sigma$-pol). For the excitation to the Ni $3d_{z^2}$ states ($\pi$-pol), the satellite feature becomes progressively broadened and less pronounced as a function of the strain tuning and oxygen content, namely, from LNO/LAO to LNO/SLAO1, and from LPNO/SLAO2 to SC LPNO/SLAO3. 

In Figs. 2a and 2b, the incident-energy dependent Ni $L_3$-RIXS spectra illustrate the overall profile of Ni $3d$ orbital excitations. Similar to the evolution of the satellite peak in the Ni $L_3$-XAS, the Ni $dd$ orbital excitations retain mostly their energy positions under the current energy resolution ($\sim$ 40 meV) while becoming gradually broadened and weakened, particularly $dd_2$ at $\sim$ 1.6 eV and $dd_3$ at $\sim$ 0.4 eV (Figs. 2c-2g), in both film series. Known from the RIXS study of the LNO bulk, $dd_3$ at $\sim$ 0.4 eV is sensitive to the interlayer $3d_{z^2}$-$2p_z$-$3d_{z^2}$  hopping~\cite{chen2024electronic}.  
The fact that the electronic structure change associated with the O $2p$ orbitals is much more significant comparing to that of the Ni $3d$ orbitals is reminiscent of a typical charge-transfer type Mott system.   

Generally, a well-defined XAS peak or a $dd$ orbital excitation is indicative of the Mott phase of a strongly correlated system; contrarily, a broadened or featureless XAS or RIXS excitation usually results from a rather delocalized system \cite{ament2011resonant}. The negligible changes in excitation energies indicate that the local crystal field is only weakly affected by the tuning. The spectral weight transfer and the broadening illustrated in the spectroscopic data highlight that the out-of-plane orbitals become delocalized under the drive of the compressive strain and the oxygenation from the insulating state towards the superconducting state.  

\begin{figure*}[]
\centerline{\includegraphics[width=120mm]{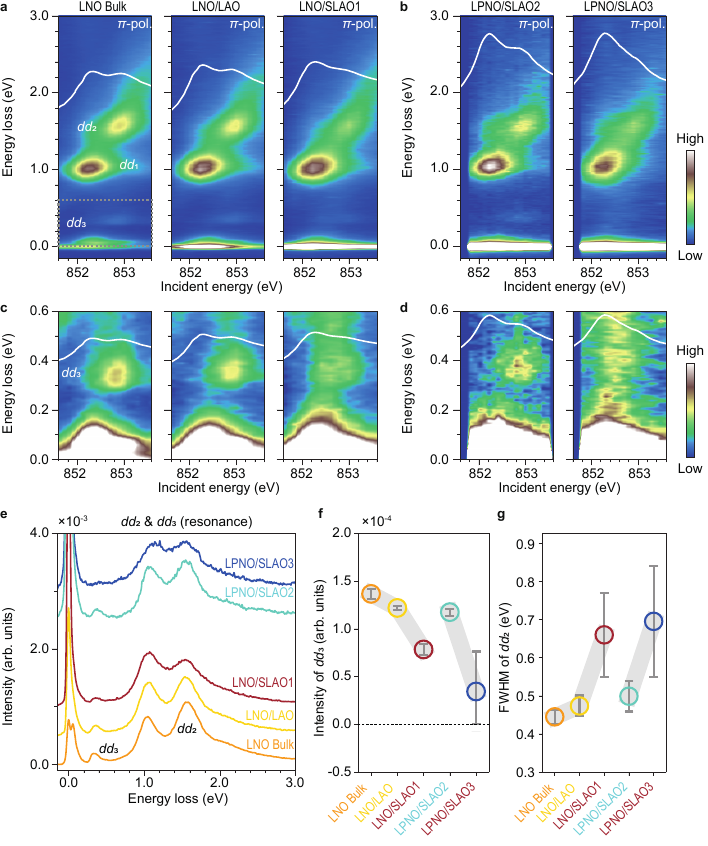}}
\caption{
\textbf{The electronic excitations in the compressively strained (La,Pr)$_3$Ni$_2$O$_{7-\delta}$ films.} \textbf{a,b,} TFY XAS and energy-dependent RIXS maps at the Ni $L_3$ edge. The RIXS spectra are normalized by the high-energy $dd$ excitations with integrated intensity over the energy-loss window [0.6 eV, 3.0 eV]. Both XAS and RIXS were measured at a grazing incidence angle of 20$^\circ$ with $\pi$ polarization to enhance sensitivity to out-of-plane $d_{z^2}$ orbital states. The RIXS map of LNO bulk is adapted from Ref.~\cite{chen2024electronic}.
\textbf{c,d,} Zoomed-in views of the RIXS maps shown in (a,b), highlighting the $dd_3$ excitations.
\textbf{e,} Integrated RIXS intensity over the incident-energy range [852.5 eV, 853.3 eV], corresponding to the resonance of the $dd_2$ and $dd_3$ excitations.
\textbf{f,} Intensity of the $dd_3$ excitation across bulk and different thin films.
\textbf{g,} FWHM of the $dd_2$ excitation across LNO bulk and thin films. Both $dd_2$ and $dd_3$ features are fitted using Lorentzian functions, with error bars representing the standard deviations of the fitting parameters. 
}
\label{fig2}
\end{figure*}

\begin{figure}[!htbp]
\centerline{\includegraphics[width=86mm,angle=0]{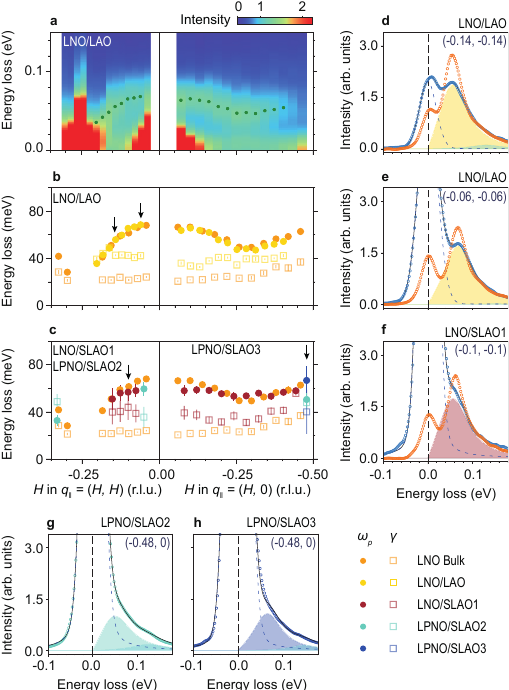}}
\caption{
\textbf{Magnetic excitations in the compressively strained (La,Pr)$_3$Ni$_2$O$_{7-\delta}$ films.}  
\textbf{a,} Momentum-resolved RIXS intensity maps of the LNO/LAO film along the $q_\parallel=(H,H)$ and $(H,0)$ directions. The Brillouin zone is defined based on the pseudo-tetragonal lattice \cite{chen2024electronic}. All spectra were measured at 20~K using $\pi$-polarized X-rays tuned to the magnon resonance (852.4~eV). Each RIXS spectrum was fitted with a pseudo-Voigt function for the elastic peak and a damped harmonic oscillator (DHO) function for the magnon excitation. The green filled circles denote the extracted magnon peak energies ($\omega_p$), with error bars representing the fitting uncertainties, which are smaller than the marker size (see details in the Supplementary Information). At $q_\parallel \approx 0$, strong elastic scattering arises from the specular geometry; these data are excluded from the colormap.
\textbf{b,c,} Magnon dispersion ($\omega_p$ vs $q_\parallel$) for LNO/LAO (yellow circles) (b); and LNO/SLAO1 (brown circles), LPNO/SLAO2 (turquoise circles), and LPNO/SLAO3 (blue circles) (c). The dispersions of LNO bulk  (orange circles) are overlaid for comparison, which were extracted from Ref.~\cite{chen2024electronic} and processed. The corresponding damping factors ($\gamma$) are shown as semi-transparent squares. Representative fitted RIXS spectra at selected $q_\parallel$ point (indicated by arrows) are shown in panels (d–h).
\textbf{d-h,} Representative RIXS spectra and corresponding fits at selected $q_\parallel$ points (indicated in each panel). The magnon spectral weight is highlighted by colored shading, and the dashed lines denote the fitted pseudo-Voigt elastic peaks. In panels (d-f), RIXS spectra of LNO bulk \cite{chen2024electronic} at the same $q_\parallel$ point (orange markers) are overlaid for comparison and their intensities were normalized to the high-energy $dd$ excitations. 
}
\label{fig3}
\end{figure}

~\\
\subsection{Magnetic excitations}

Figure 3a shows the momentum-dependent magnetic excitations in LNO/LAO, along the two high-symmetrical directions (see Methods). The damped harmonic oscillator (DHO) function was used to fit the magnons. The extracted magnon peaks ($\omega_p$) exhibit a dispersion overlapping with that of LNO bulk at ambient pressure (Fig.~3b)~\cite{chen2024electronic}. For a detailed comparison, representative RIXS spectra at $\mathbf{q}_\parallel = (-0.14, -0.14)$ (Fig.~3d) and $\mathbf{q}_\parallel = (-0.06, -0.06)$ (Fig.~3e) are presented. In contrast to the LNO bulk, the magnons in the thin films are significantly damped. The extracted damping factor ($\gamma$) is $\sim 40$~meV in the thin film twice of that in the LNO bulk.

With the similar $\omega_p$ and increased $\gamma$, the fitted undamped energy ($\omega_{\mathrm{ud}}$) differs slightly between the bulk and the thin film (within $\sim 10$~meV; see details in Supplementary information), consistent with previous reports~\cite{zhong2025spin}. In the more strongly strained non-SC LNO/SLAO1 sample, the magnon presents nearly the same dispersion as that of the LNO bulk and the thin film LNO/LAO. The extracted damping factor differs little between the LNO/SLAO1 and the LNO/LAO (Figs.~3c,3f).

For the oxygen content dependent thin film series, non-SC LPNO/SLAO2 and SC LPNO/SLAO3, despite the dominant elastic signal in RIXS spectra owing to the unit-cell thick films, magnons are recognizable as an asymmetrical tail to the elastic peak (Figs. 3g,3h). Using the DHO fitting method, the magnetic excitation and the damping factor at several momenta are extracted, as shown in Fig. 3c. Within the instrumental energy resolution, the tuning of oxygen content does not bring any discernible impact on the magnon dispersion and damping. 

Compared to the well-defined magnons in the LNO bulk which is approximate to a strongly correlated system, the much damped magnons in the thin films nearing a more itinerant electron system portray a paramagnon with reduced lifetime and coherence under either the compressive strain or the oxygen content tuning driving ultimately towards the superconducting phase. The latter is much like the paramagnons well established in cuprate superconductors \cite{le2011intense}. 


\vspace{4mm}
\subsection{Spin-density-wave}

\begin{figure}
\centerline{\includegraphics[width=86mm,angle=0]{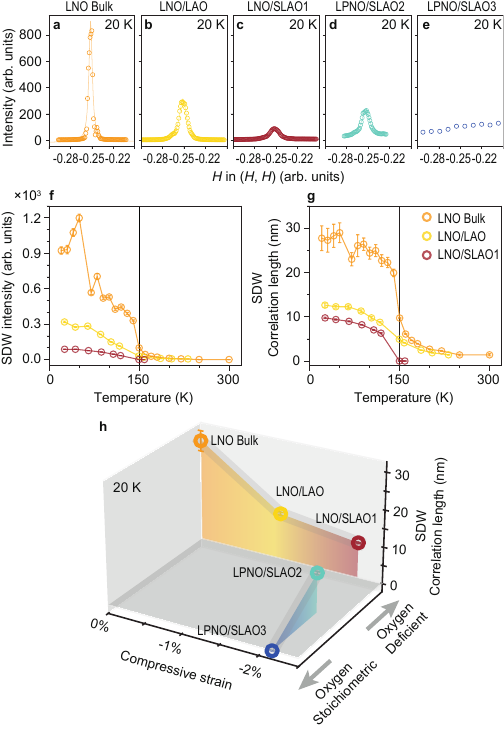}}
\caption{
\textbf{Spin-density-wave behavior in the compressively strained (La,Pr)$_3$Ni$_2$O$_{7-\delta}$ films.} 
\textbf{a–e,} SDW peaks of La$_3$Ni$_2$O$_{7-\delta}$ bulk single crystals and compressively strained thin films. Spectra were measured along the $(H,H)$ direction near the SDW wave vector $(-0.25,-0.25)$ at 20~K with $\pi$-polarized X-rays tuned to the SDW resonance (852.4~eV). Intensities were normalized to the $dd$ excitations at 1~eV, and each spectrum was fitted with a Lorentzian function.
\textbf{f,g,} Temperature dependence of the SDW intensity (f) and correlation length ($\xi_{\mathrm{SDW}}$) (g) for the bulk and thin film samples. The dashed line at 150~K indicates the SDW transition temperature of the bulk sample. The correlation length is defined as $\xi_{\mathrm{SDW}} = 1/\mathrm{HWHM}$, where HWHM is the half-width at half-maximum of the SDW peak.
\textbf{h,} Summary of the $\xi_{\mathrm{SDW}}$ as a function of compressive strain and oxygen content for the bulk and thin film samples at 20~K. All data for the bulk single crystal are extracted and processed from Ref.~\cite{chen2024electronic}.
}
\label{fig4}
\end{figure}

The LNO bulk exhibits a long-range SDW order, characterized by a prominent scattering peak at the in-plane wave vector $\mathbf{q}_{\parallel}^{\mathrm{SDW}} = (0.25, 0.25)$ (Fig.~4a)~\cite{chen2024electronic}. In the strain-dependent thin films, a similar SDW peak is observed at the same $\mathbf{q}_{\parallel}^{\mathrm{SDW}}$ in the low temperature (Figs.~4b,4c). Interestingly, the SDW intensity is progressively reduced upon the compressive strain tuning (Figs.~4a-4c). The SDW correlation length ($\xi$) also decreases gradually, from $\sim$ 30 nm in the bulk crystal, to $\sim$ 13 nm and $\sim$ 10 nm in LNO/LAO and LNO/SLAO1, respectively (Fig. 4g). Furthermore, both the SDW intensity and the correlation length manifest clear temperature dependence: they weaken gradually at elevated temperatures and becomes barely discernible above $\sim$ 150 K (Figs.~4f,g). In the non-SC LPNO/SLAO2, the SDW order is present at the wavevector $\mathbf{q}_{\parallel}^{\mathrm{SDW}}$. The SDW intensity and correlation length are characterized with comparable values as that in the strain-tuned films. Remarkably, following a thorough investigation, no SDW order was revealed in the SC LPNO/SLAO3 film (Fig.~4e) (see Supplementary information). Since superconductivity was ensured prior and after the RIXS measurement, the non-discernible SDW signal confirms that the long-range magnetic order competes with superconductivity and is fully suppressed in the superconducting state. 

If the LNO bulk is viewed as the ‘undoped' parent compound, the strained non-SC LNO/LAO, LNO/SLAO1, and the oxygen deficient non-SC LPNO/SLAO2 films can be regarded as ‘underdoped' LNO. Here, the reduced SDW intensity and correlation length are signatures of competition between magnetism and superconductivity instability before it gets fully suppressed in the near `optimally' doped SC LPNO/SLAO3 sample (Fig. 4h). Dissecting the phase-diagram like picture informs us that the oxygen vacancies are detrimental to superconductivity~\cite{liu2023s,lu2025impact,wang2025electronic2,jiang2025dual}, and, the eventual realization of bulk superconductivity in La$_3$Ni$_2$O$_{7-\delta}$ films requires both the compressive strain and the macroscopic oxygen stoichiometry~\cite{ko2025signatures,zhou2025ambient,SLAO60K,tarn2026reducing}.

\section{Discussion}

The combination of X-ray absorption spectroscopy and the resonant inelastic X-ray scattering unfolded a rich landscape of the electronic structure and the magnetic properties of the (La,Pr)$_3$Ni$_2$O$_{7-\delta}$ thin films under the two parallel tuning routes $i.e.$, the compressive strain and the oxygen content. we firstly learned from the O $K$-XAS that the spectral weight transfer from the upper Hubbard band like peak to the O $2p_z$ hole state is realized upon tuning. The broadening effect encoded in the Ni $L$-XAS and RIXS collectively confirms that the tuning drives the delocalization of the correlated $d_{z^2}$ state. Both the spectral weight transfer and the delocalization are well-known spectroscopic fingerprints imprinted in the hole-doped cuprates evolving from an antiferromagnetic insulator to a more delocalized superconductor~\cite{chen1991electronic,eskes1991anomalous}. It is remarkable to see the above spectroscopic fingerprints are concerted in the two independent tuning routes in (La,Pr)$_3$Ni$_2$O$_7$ when driving towards the superconducting state. 

Uncovering the development of the electronic structure of the RP nickelates from the non-superconducting to the superconducting phase is one of the salient results of our study which can provide a strong constraint for future theoretical studies. Importantly, the predominant involvement of the O $p_z$ and the Ni $3d_{z^2}$ orbitals in the process lends to a picture that the interlayer $3d_{z^2}$-$2p_z$-$3d_{z^2}$ hybridized molecular orbital channel needs to become delocalized before getting involved in superconductivity~\cite{sun2023signatures}. 
The negligible involvement of the Ni $3d_{x^2-y^2}$ orbital in the tuning process is consistent with the fact that it is already itinerant prior to forming the Cooper pairing~\cite{yang2024orbital}. Also, the fixed $dd$ orbital excitation energies as well as the near constant absorption thresholds of the O $K$- and Ni $L_3$- edges allude to a robust crystal-field splitting with respect to the Fermi level. Our result differs from the DFT calculations, which suggest a substantial shift of the $d_{z^2}$-dominated band away from the Fermi level (by several hundred meV)~\cite{bhatta2025structural,geisler2025fermi}.   

The short-range spin fluctuation is regarded as one of the potential mechanisms for the unconventional superconductivity. Previously the dispersive magnon bandwidth in La$_3$Ni$_2$O$_7$ was reported to increase under the compressively-strained films~\cite{zhong2025spin}. Our RIXS results reveal that the magnon becomes damped upon tuning of the strain or the oxygen content without altering its bandwidth or the magnetic energy scale. The damped magnon may be due to the increased itinerancy and the robust magnon bandwidth is reconcilable with the unchanged crystal field splitting. However, our result is in stark contrast to the theoretical proposal of a significant enhancement of the interlayer exchange coupling ($J_{\perp}$) by nearly 50\% in the superconducting phase compared to under the ambient-pressure~\cite{yi2025unifying}. Despite the robust short-range magnetic fluctuation, the long-range SDW order shows a strong reduction in both ordering intensity and the correlation length upon the tuning of the strain and the oxygen content, establishing its direct competition with superconductivity consistent with the findings by the $\mu$SR and optical studies~\cite{khasanov2025pressure,meng2024density}. The result advocates a close link between RP nickelates and other unconventional high-temperature superconductors.

In summary, the study of (La,Pr)$_3$Ni$_2$O$_{7-\delta}$ thin films reveals that compressive strain and oxygenation collectively drive the delocalization of the $3d_{z^2}$ and $2p_z$ orbitals and the suppression of long-range spin-density-wave order, which together induce the high-temperature superconducting state. This scenario is evidenced by three spectroscopic signatures: (1) spectral weight transfer in XAS, (2) broadening of the $dd$ excitations, and (3) damping of magnon excitations. These findings establish the essential prerequisites for superconductivity in the Ruddlesden-Popper nickelates, paving a definitive path for the design of novel superconductors.




\textit{Note added:} Upon completion of this work, we became aware of an independent RIXS study on superconducting (La,Pr)$_3$Ni$_2$O$_7$ thin films \cite{zhang2026interlayer}.

\section{Methods}
\noindent\textbf{Growth of thin films:} For the
strain-tuning thin films (non-SC LNO/LAO and non-SC LNO/SLAO1). High-quality La$_3$Ni$_2$O$_{7-\delta}$ thin films were deposited on (001)-oriented LaAlO$_3$ and SrLaAlO$_4$ substrates with a uniform thickness of approximately 30~nm using pulsed laser deposition (PLD). A stoichiometric La$_3$Ni$_2$O$_7$ polycrystalline target was synthesized via a solid-state reaction route employing La$_2$O$_3$ and NiO powders. Prior to deposition, the base pressure was maintained at 1.33 $\times$ 10$^{-8}$ mbar. During the growth process, the oxygen partial pressure and substrate temperature were set at 4 $\times$ 10$^{-1}$ mbar and 750 $^\circ$C, respectively. The energy density was 1.25 J cm$^{-2}$, and the laser frequency was maintained at 5 Hz. All films underwent post-annealing in an oxygen-rich atmosphere ($P_{\mathrm{O}_2}$ = 1.33 $\times$ 10$^2$ mbar), followed by a slow cooling to room temperature. The structural properties of the films were characterized using X-ray diffraction techniques, including 2$\theta$-$\omega$ scans, rocking curves, and reciprocal space mapping (RSM), with a Panalytical X’Pert3 MRD diffractometer equipped with a Cu $K_{\alpha1}$ source. The film thickness was determined through X-ray reflectivity (XRR) and analyzed using GenX software. Cross-sectional transmission electron microscopy (TEM) specimens, exhibiting various strain states, were prepared via the focused ion beam (FIB) lift-off process. High-angle annular dark-field scanning transmission electron microscopy (HAADF-STEM) and annular bright-field scanning transmission electron microscopy (ABF-STEM) experiments were conducted using a JEM ARM 200CF microscope at the Institute of Physics (IOP), Chinese Academy of Sciences. ~\\

For the oxygen-content-tuning thin films (non-SC LPNO/SLAO2 and SC LPNO/SLAO3). High-quality (La,Pr)$_3$Ni$_2$O$_{7-\delta}$ films were grown using the gigantic-oxidative atomic-layer-by-layer epitaxy (GAE) method \cite{Zhou2025gigantic,SLAO60K}, in which (La,Pr)O$_x$ and NiO$_x$ targets were alternately ablated using a pulsed laser. The (La,Pr)O$_x$ target has a nominal composition of (La$_{0.65}$Pr$_{0.35}$)O. This alternating deposition sequence produces a stacking structure of (La,Pr)O–NiO$_2$–(La,Pr)O–NiO$_2$–(La,Pr)O, resulting in a nominal composition of La$_{1.95}$Pr$_{1.05}$Ni$_2$O$_{7-\delta}$. On as-received SrLaAlO$_4$ substrates (MTI-Kejing), a (La,Pr)O–NiO$_2$–(La,Pr)O buffer layer was first deposited to mitigate interfacial structural discontinuities. During growth, layer-by-layer control was achieved by synergistically tuning the laser pulse energy and pulse sequence, together with real-time monitoring of the reflection high-energy electron diffraction (RHEED) intensity oscillations. No post-annealing treatment was performed for any of the samples. 

The non-SC LPNO/SLAO2 and SC LPNO/SLAO3 were grown at a substrate temperature of 850~$^\circ$C, measured from the back side of the Inconel flag-type sample holder. For SC LPNO/SLAO3, the laser pulse ratio was 101:148 under an ozone partial pressure of $1.04\times10^{-2}$ mbar and the same total oxygen pressure of $1.15\times10^{-1}$ mbar. After film growth, the sample was first cooled at 100~$^\circ$C/min to below 200~$^\circ$C under the same oxidative environment. Subsequently, an SrTiO$_3$ capping layer was deposited at room temperature using 800 laser pulses at 8 Hz and 600 mJ laser energy under the same growth atmosphere. For non-SC LPNO/SLAO2, the laser pulse ratio was set to 107:138 (pulses on (La,Pr)O$_x$ : NiO$_x$). The film was grown in a mixed ozone/oxygen atmosphere with an ozone partial pressure of $1.40\times10^{-2}$ mbar and a total oxygen pressure of $1.15\times10^{-1}$ mbar. After deposition, a SrTiO$_3$ capping layer was grown $in~situ$ using 400 laser pulses at 4 Hz with a laser energy of 600 mJ under the same oxidative conditions while maintaining the substrate temperature at 850~$^\circ$C. The sample was then cooled at a rate of 100~$^\circ$C/min to below 200~$^\circ$C in the same atmosphere. ~\\

\noindent\textbf{XAS and RIXS measurements:} Experiments were performed at Beamline I21 at Diamond Light Source~\cite{zhou2022i21}. In this work, we describe the structural properties of (La,Pr)$_3$Ni$_2$O$_{7-\delta}$ thin films referencing a pseudo-tetragonal unit cell with lattice constants $a^T$, $b^T$ and $c^T$ \cite{chen2024electronic}. For direct comparison among different samples, reciprocal lattice units (r.l.u.) are defined using the lattice parameters of La$_3$Ni$_2$O$_7$ bulk single crystal, with $a^T$ = $b^T$ $\sim$ 3.833 Å and $c$ = 20.45 Å. Accordingly, under $2\pi/a^T = 2\pi/b^T = 2\pi/c = 1$, the momentum transfer is expressed as $\mathbf{Q} = H\mathbf{a}^{T*} + K\mathbf{b}^{T*} + L\mathbf{c}^*$. The crystallographic $a^T$–$c$ ($b^T$–$c$) plane of La$_3$Ni$_2$O$_7$ single crystal was aligned within the horizontal scattering plane. Sample's polar angular offsets ($\theta$ and $\chi$) and azimuthal offset ($\phi$) were aligned by SDW order peak, such that the $c^*$ axis lies in the scattering plane. The spectrometer arm was at a fixed position of $\Omega=154^\circ$. XAS spectra were collected with a grazing incidence angle of $\theta_0 = 20^\circ$ to probe both in-plane and out-of-plane unoccupied states. All XAS measurements were done at a temperature of 20 K with the exit slit opening to 30 $\mu$m. Total fluorescence yield XAS spectra were collected and normalised to the incoming beam intensity. Both linear vertical ($\sigma$) and horizontal ($\pi$) polarisations were used. Energy-dependent RIXS measurements were performed at the grazing incidence angle of $\theta_0 = 20^\circ$ and the temperature of 20 K. The exit slit was open to 30 $\mu$m corresponding to an average energy resolution of 42 meV (FWHM). The incident energy range went from 851.6 to 853.6 eV in steps of 0.2 eV (or 0.15 eV) to fully capture the resonance behaviour across the Ni-$L_3$ absorption peaks. Momentum-dependent RIXS measurements were performed at the resonant energy of 852.4 eV at a temperature of 20 K with the exit slit opening to 20 $\mu$m corresponding to an average energy resolution of 39 meV (FWHM). The momentum resolution is 0.002 r.l.u. near the SDW wavevector at the Ni $L_3$-edge. RIXS spectra were collected using both $\pi$ polarisations. The grazing incidence geometry ($\theta < \Omega/2$) was applied for the acquisition of RIXS spectra shown in the main text.

~\\

\noindent\textbf{\large Data availability}~\\
All data generated or analyzed during this study are available from the corresponding authors upon reasonable request.
~\\


%


~\\


\noindent\textbf{\large Competing interests}~\\The authors declare no competing interests.

\end{document}